\documentstyle[pre,aps,twocolumn,psfig]{revtex}     

\newcommand{\figutwo}[4]{
                         \nobreak \hbox to \textwidth { \leavevmode 
                         \epsfxsize=#1 \epsffile{#2}
                         \hss \leavevmode 
                         \epsfxsize=#3 \epsffile{#4} }
                         \nobreak
                        }
\newcommand{\figu}[2]{\nobreak \hbox to \textwidth {
                      \centerline{\leavevmode
                      \epsfxsize=#1 \epsffile{#2}
                      }} \nobreak 
                     }

\newcommand{\bge}{\begin{equation}}
\newcommand{\ede}{\end{equation}}
\newcommand{\ba}{\begin{array}}
\newcommand{\ea}{\end{array}}

\newcommand{\h}{\hbar}

\newcommand{\f}{\frac}

\newcommand{\p}{\partial}
\newcommand{\no}{\nonumber}
\newcommand{\e}{\mbox{e}}

\input psfig.sty

\newcommand{\rf}     [1] {~\cite{#1}}

\newcommand{\beq}{\begin{equation}}
\newcommand{\continue}{\nonumber \\ }

\newcommand{\eeq}{\end{equation}}
\newcommand{\ee}[1] {\label{#1} \end{equation}}
\newcommand{\bea}{\begin{eqnarray}}

\newcommand{\eea}{\end{eqnarray}}
\newcommand{\barr}{\begin{array}}
\newcommand{\earr}{\end{array}}

\newcommand{\evOper}{evolution oper\-ator}

\newcommand{\fd}{spec\-tral det\-er\-min\-ant}

\renewcommand{\det}{\mbox{\rm det}}

\newcommand{\Lop}{{\cal L}}	   

\begin{document}
\draft{
\title{Trace formula for noise corrections to trace formulas}

\author{Gergely Palla and G\'abor Vattay}
\address{Department of Physics of Complex Systems, E{\"o}tv{\"o}s University\\
P\'azm\'any P{\'e}ter s{\'e}tany 1/A,
H-1117 Budapest, Hungary   } 

\author{ Andr{\'e} Voros} 
\address{
CEA, Service de Physique Th{\'e}orique de Saclay\\
F-91191 Gif-sur-Yvette CEDEX (France)\\
}

\date{\today}

\maketitle

\begin{abstract}
We consider an evolution operator for a discrete Langevin equation
 with a strongly hyperbolic 
classical dynamics and Gaussian noise. Using an integral
 representation of the evolution operator $\Lop$ we investigate the
 high order corrections to the  trace of $\Lop^n$. The asymptotic 
behaviour is found to be controlled by sub-dominant saddle points 
previously neglected in the perturbative expansion. We show that
 a trace formula can be derived to describe the high order
noise corrections.

\end{abstract}

\vspace{1cm}
In the statistical theory of dynamical systems the development of the
densities of particles is governed by a corresponding evolution
 operator.
For a repeller, the leading eigenvalue of this operator $\Lop$ 
yields a physically measurable property of the dynamical system,
the escape rate from the repeller.
In the case of deterministic flows, the periodic orbit theory\rf{A1,A2}
yields explicit and numerically efficient formulas for the spectrum of $\Lop$ as zeros
of its {\fd}\rf{QCcourse}.

On all dynamical evolutions in nature stochastic 
processes of various strength have an influence.
In a series of papers\rf{noisy_Fred,conjug_Fred,diag_Fred,asymp}
 the effects of noise on measurable
properties such as dynamical averages in classical
 chaotic dynamical
systems were systematically accounted.
The theory developed is  closely related to the semi-classical $\h$ expansions
\rf{Gaspard,Vattay1,Vattay2}
 based
on Gutzwiller's formula for the trace in terms of classical periodic
orbits\rf{gutbook} in that both are perturbative theories in the noise
strength or $\hbar$, derived from saddle-point expansions of a path
integral containing a dense set of unstable stationary points.
  The analogy with quantum mechanics and
field theory is made explicit in\rf{noisy_Fred} where
Feynman diagrams are used to find the lowest nontrivial
 noise corrections
 to the escape rate.

In\rf{diag_Fred}  we developed an explicit matrix
representation of the stochastic evolution operator. 
The numerical implementation made it possible to reach  up to 
order eight in expansion order, and the  corrections to the escape 
rate were found to be a divergent series in the 
noise expansion parameter. This reflects  that the corrections were 
calculated (using the
so called cumulant expansion) from other divergent quantities, the 
traces of the evolution operator $\Lop^n$ \rf{diag_Fred}.

In\rf{asymp} the focus was on  the high order noise
corrections for the special case of the first trace,  Tr$\Lop$. 
The asymptotics of  the trace of 
the evolution operator were governed by  sub-dominant saddles 
previously neglected in the expansion.

In this paper we  show that the high order noise corrections of 
 Tr${\cal L}^n$ are also dominated by sub-dominant saddles.
These sub-dominant saddles can be treated as generalised periodic orbits of
 the system and we associate them with periodic orbits of  corresponding 
 discrete Newtonian equations of motion. Our key result is ($\ref{result2}$)
 where 
the high order noise corrections are converted into a trace formula. 
We give as a numerical example the quartic map considered
 in\rf{noisy_Fred,conjug_Fred,diag_Fred,asymp}.

First we introduce the noisy repeller and its
 evolution operator. 
An individual trajectory in presence of additive noise is generated
by iterating 
\beq
x_{n+1}=f(x_{n})+\sigma\xi_{n} 
\,,
\ee{mapf(x)-Diag}
where $f(x)$ is a map, 
$\xi_n$ a random variable with the normalised
distribution $p(\xi)$, 
and $\sigma$ parametrises the noise strength.
In what follows we shall assume that the mapping $f(x)$ is
one-dimensional and expanding, and that the $\xi_n$ are uncorrelated.
A density of trajectories $\phi(x)$ evolves with time on the average as
\beq
\phi_{n+1}(y) =
\left(
\Lop
\circ
\phi_{n}\right)(y)
= \int dx \, \Lop(y,x) \phi_{n}(x)
\ee{DensEvol}
where the $\Lop$ {\evOper} has the general form 
\bea
\Lop(y,x) &=& \delta_\sigma(y-f(x)) ,\label{OpOverNoise}
        \continue
  \delta_\sigma(x)
  &=& \int \delta(x-\sigma \xi) p(\xi) d\xi 
  \,=\, \frac{1}{\sigma} p\left( \frac{x}{\sigma} \right)
\,.
\label{oper-Diag}
\eea
 For the calculations
in this paper Gaussian weak noise is assumed. In the perturbative
 limit, $\sigma \rightarrow 0$, the evolution operator becomes
\bea
\Lop(x,y)&=&\f{1}{\sqrt{2\pi}\sigma}{\e}^{-\f{(y-f(x))^2}{2\sigma^2}}.   
\eea
The map considered here is the same as in our previous papers, a
 quartic map on the
$(0,1)$ interval given by
\bea
f(x)=20\left[\f{1}{16}-\left(\f{1}{2}-x\right)^4\right].
\eea
 Throughout the theory developed in previous 
works\rf{noisy_Fred,conjug_Fred,diag_Fred,asymp} , 
the periodic orbits of the system played a major role. A 
periodic orbit of length $n$ was defined simply by
\bea
x_{j+1}&=&f(x_j),\mbox{\hspace{1cm}}j=1,...,n\label{peri1}\\
x_{n+1}&=&x_{1}. \label{peri2}
\eea

For a repeller the leading eigenvalue of the evolution operator
yields a physically measurable property of the dynamical system,
the escape rate from the repeller.
In the case of deterministic flows, the periodic orbit theory
yields explicit formulas for the spectrum of ${\cal L}$ as zeros
of its spectral determinant \cite{QCcourse}. One of the most important
goals of the theory related to stochastic evolution operators  is to
 explore the dependence of the eigenvalues 
$\nu$ of ${\cal L}$ 
on the noise strength parameter $\sigma$.
The eigenvalues are determined by the eigenvalue condition
\bea
F(\sigma,\nu(\sigma))=\det(1-{\cal L}/\nu(\sigma)) =0
\label{eigCond}
\eea 
where
$
F(\sigma,1/z) = \det(1-z{\cal L})
$
is the spectral determinant of  the evolution operator $\Lop$, which
can be expressed as
\bea
         \det(1-z\Lop)
        =
        \exp\left(-\sum_n^\infty {z^n \over n} \mbox{Tr} \Lop^n \right)
\,.
\label{det-tr-Diag}
\eea
Equation ($\ref{det-tr-Diag}$) shows that noise dependence of the
 eigenvalues of the evolution operator are very closely related to the
 noise dependence of the trace of $\Lop^n$, which shall be the object
 of study from now on.

The trace of $\Lop^n$ can be expressed as
\bea
\mbox{Tr}{\cal L}^n=\f{1}{(\sqrt{2\pi}\sigma)^n}\int dx_1dx_2...dx_n
{\e}^{-\f{S}{\sigma^2}},
\label{trace}
\eea
where
\bea
S&=&\f{1}{2}\sum_{j=1}^n\left(x_{j+1}-f(x_j)\right)^2, \label{Sdef}\\
x_{n+1}&=&x_1.
\eea

 In order to give a deeper insight on the forthcoming calculations
we draw a correspondence between discrete Hamiltonian mechanics and our system,
with the $S$ defined above playing the role of the classical action.
According to ($\ref{Sdef}$),
 the least action principle requires
\bea
x_j-f(x_{j-1})-f'(x_j)(x_{j+1}-f(x_j))=0.
\label{least}
\eea
We define 
\bea
p_j:=x_j-f(x_{j-1}),
\eea
the quantity corresponding to the momentum in the classical mechanics.
From (\ref{least}) we obtain
\bea
x_{j+1}&=&f(x_j)+p_{j+1}, \\
p_{j+1}&=&\f{p_j}{f'(x_j)},
\eea
which are the equations corresponding to the classical Newtonian equations
 of motion.
The generalised periodic orbits of length $ n$ are those orbits, 
which obey these equations and $x_{n+1}=x_1$,
$p_{n+1}=p_n$. Those generalised periodic orbits which have
non-zero momentum will control the asymptotic behaviour
of the corrections to $\mbox{Tr}{\cal L}^n$ 
as we shall demonstrate later.
The original periodic orbits defined by ($\ref{peri1}$),($\ref{peri2}$)
are those with zero momentum. 
 The generalised periodic orbits
with non-zero momentum and the original periodic orbits proliferate
 with growing $n$ as suggested by Fig $\ref{orbits}$.
\begin{figure}[hbt]
\centerline{\strut\psfig{figure=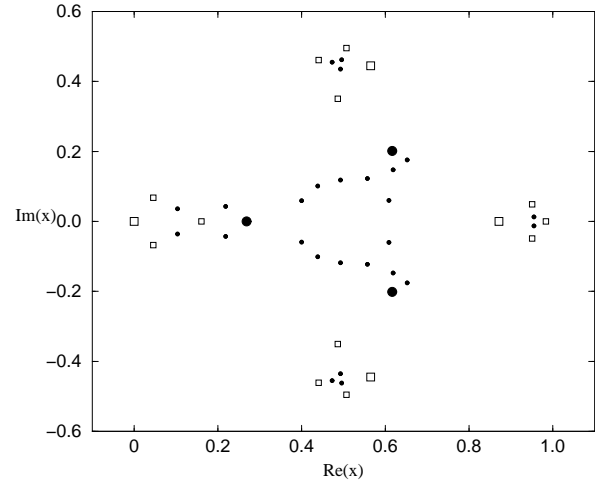,height=6.5cm}}
\caption{ The sets of original and generalised periodic orbits. Squares
 indicate original periodic orbits, dots indicate generalised periodic orbits,
 large symbols indicate orbits of length one, small symbols indicate orbits
 of length two.}
\label{orbits}
\end{figure}

We introduce an integral representation of the 
noisy kernel, which will be of great use in the later
 calculations:
\bea
\Lop(x,y)&=&\f{1}{\sqrt{2\pi}\sigma}{\e}^{-\f{(y-f(x))^2}{2\sigma^2}}=
 \no \\ & &
\f{1}{2\pi}\int dk {\e}^{-\f{\sigma^2k^2}{2}+ik(y-f(x))}. \label{theop}
\eea
Using  this new integral representation,
\bea
& &\mbox{Tr}{\cal L}^n=\no \\ 
& &\f{1}{(2\pi)^n}\int dk^ndx^n {\e}^{-\f{\sigma^2}{2}
\sum_{j=1}^nk_j^2+i\sum_{j=1}^nk_j(x_{j+1}-f(x_j))},\no \\
\label{ujtrace}
\eea
or equivalently
\bea
& &\mbox{Tr}{\cal L}^n=\no \\ 
& &\f{1}{(2\pi)^n}\int dk^n \int dp^n J(p){\e}^{-\f{\sigma^2}{2}
\sum_{j=1}^nk_j^2+i\sum_{j=1}^nk_jp_j}, \label{moment}
\eea
where $J(p)=D(x)/D(p)$ denotes the Jacobian. 
Since
\bea
\f{1}{(2\pi)^n}\int dk^n {\e}^{i\sum_{j=1}^n k_jp_j}=\prod_{j=1}^n
\delta(p_j),
\eea
we can reduce ($\ref{moment}$) to 
\bea
\mbox{Tr}{\cal L}^n&=&\int dp^n J(p){\e}^{\f{\sigma^2}{2}\Delta}
\prod_{j=1}^n\delta(p_j)=\no \\ & &
\left.{\e}^{\f{\sigma^2}{2}\Delta}J(p)
\right|_{p_j=0}, \label{elegant}
\eea
where $\Delta$ denotes the Laplacian
\bea
\Delta=\f{\p^2}{\p p_1^2}+\f{\p^2}{\p p_2^2}+...+\f{\p^2}{\p_n^2}.
\eea
Our object of study is the Taylor expansion of ($\ref{elegant}$) in the
noise parameter:
\bea
\mbox{Tr}{\cal L}^n&=&\sum_{N=0}^{\infty}\left(\mbox{Tr}{\cal L}^n\right)_N
\sigma^{2N},\\
\left(\mbox{Tr}{\cal L}^n\right)_N&=&
\left.\f{1}{2^N}\f{\Delta^N}{N!}J(p)\right|_{p_j=0}.
\eea
The $N$-th power of the Laplacian in the equation above can be written as
\bea
\Delta^N=
\sum_{j_1,...,j_n=0}^{\infty}\f{N!}{j_1!...j_n!}
\f{\p^{2j_1}}{\p p_1^{2j_1}}...\f{\p^{2j_n}}
{\p p_n^{2j_n}}\delta_{N,\sum_{k=1}^nj_k}, 
\eea
where $\delta_{jl}$ is the Kronecker-delta.
With the help of the multidimensional residue formula from complex calculus
 \rf{comcalc} 
\bea
& &\f{\p^{n_1+...+n_k}f(z)}{\p z_1^{n_1}...\p z_k^{n_k}}=\no \\ & &
\f{n_1!...n_k!}{(2\pi {i})^k}\oint_{c_1}...\oint_{c_k}\f{
f(\xi)d\xi_1...d\xi_k}{(\xi_1-z_1)^{n_1+1}...(\xi_k-z_k)^{n_k+1}},
\eea
we obtain
\bea
& &\left(\mbox{Tr}{\cal L}^n\right)_N=\f{1}{(2\pi {i})^n2^N}
\sum_{j_1,...,j_n=0}^{\infty}\f{(2j_1)!}{j_1!}...
\f{(2j_{n}))!}{j_{n}!}\no \\ & &\times\delta_{N,\sum_{k=1}^nj_k}
\oint_{c_1}...
\oint_{c_n}\f{J(p)dp_1...dp_n}{p_1^{2j_1+1}...
p_n^{2j_n+1}}.\label{numeri1}
\eea
The contours are around the $p_j=0$ points. 
The integrals can be transformed back to contour integrals in 
the original $x_j$ variables, and the contours will be placed
 around the original periodic orbits of the system defined
by ($\ref{peri1}-\ref{peri2}$), since it is these orbits
 which fulfil the $p_j=0$ conditions. From now on we shall restrict
 our calculations to the asymptotic large $N$ limit. We will replace
 the summations in ($\ref{numeri1}$) by integrals and then use the
 saddle-point method to get a compact formula for 
$(\mbox{Tr}{\cal L}^n)_N$.
We approximate the factorials via the Stirling-formula \rf{abramo} as
\bea
\f{(2j_k)!}{j_k!}&\simeq&\f{\left(\f{2j_k}{e}\right)^{2j_k}\sqrt{4\pi j_k}}
{\left(\f{j_k}{e}\right)^{j_k}\sqrt{2\pi j_k}}=2^{2j_k+1/2}j_k^{j_k}
{\e}^{-j_k}=\no \\ & &2^{1/2}{\e}^{2(\ln 2)j_k+j_k\ln j_k-j_k}. 
\label{StirlingII}
\eea
Using ($\ref{StirlingII}$) and an integral representation of the
 delta function we get
\bea
& &\left(\mbox{Tr}{\cal L}^n\right)_N\simeq\f{2^{\f{n}{2}-N}}{(2\pi {i})^n2\pi}
\sum_{j_1,...,j_n=0}^{\infty}\int dt \oint_{c_1}...\oint_{c_n}dx_1...dx_n
\no \\ & &
\times\exp\left[{i}t(N-\sum_{k=1}^nj_k)+(2\ln 2-1)\sum_{k=1}^nj_k
+\sum_{k=1}^nj_k\ln j_k\right.\no \\& &\left.+\sum_{k=1}^n\ln(x_k-f(x_{k-1}))
(2j_k+1)\right].
\eea
Now we replace $j_k$ with the new variables $y_k=\f{j_k}{N}$ and 
in the asymptotic ($N$ large) limit approximate the summations by $y_k$
 with integrals by $y_k$ as
\bea
& &\left(\mbox{Tr}{\cal L}^n\right)_N\simeq\no \\ & &
\f{2^{\f{n}{2}-N}N^n}{(2\pi {i})^n2\pi}
\int_0^{\infty}dy_1...\int_0^{\infty}dy_n
\int dt \oint_{c_1}...\oint_{c_n}dx_1...dx_n
\no \\ & &
\times\exp\left[{i}t(N-N\sum_{k=1}^ny_k)+N(2\ln 2-1)\sum_{k=1}^ny_k
\right.\no \\ & &\left.+N\sum_{k=1}^ny_k\ln(Ny_k)
+\sum_{k=1}^n\ln(x_k-f(x_{k-1}))(2Ny_k+1)\right].\no \\
\eea
 We evaluate the $y$ integrals with the saddle point
 method to get 
\bea
\left(\mbox{Tr}{\cal L}^n\right)_N&\simeq&\f{2^{-N+\f{n}{2}}}
{(2\pi)^{\f{n}{2}}{i}^n2\pi}
\int dt \oint_{c_1}...\oint_{c_n}
dx_1...dx_n\no \\ & &
\exp\left[{i}t\left(N+\f{n}{2}\right)-{\e}^{it}\f{S}{2}\right].
\label{prefactor}
\eea
Next we implement the saddle point method to the
integral in $t$ as well, asymptotically resulting in
\bea
\left(\mbox{Tr}{\cal L}^n\right)_N&\simeq&
\f{N^{\f{n-1}{2}}}{2^{2N+\f{1}{2}}(2\pi)^{\f{n+1}{2}}{i}^{n+1}}
\f{(2N)!}{N!}\no \\ & &\times
\int dx^n{\e}^{-\left(N+\f{n}{2}\right)\log(S)}. \label{ucso}
\eea 
The last step is to evaluate the contour integrals in the $x_k$
variables. We deform the contours, until the saddle points are reached
so the contours run along the routes of the steepest descent. The
 leading contribution comes from the saddle points, which fulfil
 the following equation
\bea
\f{1}{S}\left[
x^*_{j}-f(x^*_{j-1})-(x^*_{j+1}-f(x^*_j))f'(x^*_j)\right]=0.
\label{saddle2}
\eea
By comparing ($\ref{saddle2}$) and ($\ref{least}$) one can see that
the saddle-points are all generalised periodic orbits of the system.
 Since the contours ran originally around the 
 orbits with zero momentum, these do not come into account as saddle points.
The second derivative matrix is
\bea
-\left(N+\f{n}{2}\right)\f{1}{S}D^2 S,
\eea
where $D^2 S$ denotes the second derivative matrix of $S$
\bea
(D^2S)_{ij}=\f{\partial^2S}{\partial x_i\partial x_j}.
\eea
This would be the matrix to deal with if we would have taken the 
saddle point approximation of ($\ref{trace}$) directly.
We reorganise the prefactor in ($\ref{ucso}$) with the use of the 
Stirling formula \rf{abramo} and the result of the saddle point integration
is written as
\bea
& &(\mbox{Tr}{\cal L}^n)_N\simeq\no \\ & &
\sum_{s.p.}\f{N^{\f{n-1}{2}}}{2\pi {i}}
\f{\Gamma(N+\f{1}{2})}{\left(N+\f{n}{2}\right)^{\f{n}{2}}}
\f{S_p^{-N}}{\sqrt{\det D^2 S_p}}, \label{result}
\eea
which is our main result. For $n=1$ this formula gives back the
 result of \rf{asymp} as it should.

Finally we draw the attention to the
 close connection between  the generalised periodic orbits of the 
 system and $D^2S$.
The stability matrix of a general periodic orbit is expressed
 as
\bea
J&=&J_1\cdot J_2\cdot J_3...\cdot J_n \\
J_k&=&\left(\matrix{f'(x_k)-\f{p_k}{(f'(x_k))^2}f''(x_k) & 
\f{1}{f'(x_k)} \cr -\f{p_k}{(f'(x_k))^2}f''(x_k) & \f{1}{f'(x_k)}}
\right)
\eea
The determinant of $D^2S$ can be expressed with the help of
 the stability matrix as
\bea
\det D^2S_p=\det(J_p-1).
\eea
This way we reformulate ($\ref{result}$) as
\bea
(\mbox{Tr}{\cal L}^n)_N\simeq\f{N^{\f{n-1}{2}}}{2\pi}
\f{\Gamma(N+\f{1}{2})}{\left(N+\f{n}{2}\right)^{\f{n}{2}}}
\sum_{p.o.}\f{{\e}^{-N\log S_p}}{\sqrt{\det(1-J_p)}}, \label{result2}
\eea
where the summation runs over generalised periodic orbits, with
  non-zero momentum. This is  fully  analogous to a trace formula and
 is our main result. 

Finally we turn towards testing our result obtained so far.
In \rf{asymp} we developed a contour integral method to calculate
 high order noise corrections to the trace of $\Lop$. We 
 showed that the agreement between the exact results and a formula
 which coincides with the ($\ref{result}$) in the $n=1$ case is very
 good. Now we step ahead and produce numerically high order noise
corrections to the trace of $\Lop^2$. We shall start from 
($\ref{numeri1}$) by transforming the integrals in $p$ back to
 integrals in $x$ as
\bea
& &\left(\mbox{Tr}{\cal L}^n\right)_N=\f{1}{(2\pi {i})^n2^N}
\sum_{j_1,...,j_n=0}^{\infty}\f{(2j_1)!}{j_1!}...
\f{(2j_{n}))!}{j_{n}!}\no \\ & &\times\delta_{N,\sum_{k=1}^nj_k}
\oint_{c_1}...
\oint_{c_n}\no \\ & &\f{dx_1...dx_n}{(x_1-f(x_n))^{2j_1+1}...
(x_n-f(x_{n-1}))^{2j_n+1}}.
\eea
The contours at ($\ref{numeri1}$) were around the $p_j=0$ points, so
the contours above are placed around the original periodic orbits
 of the system, defined by ($\ref{peri1}$),($\ref{peri2}$). 
These contour integrals
can be  evaluated numerically. The
 Fig. $\ref{abra}$ shows the ratio of  
 $(\mbox{Tr}\Lop^2)_N$ obtained from ($\ref{result}$)
 and evaluated via the procedure described above as a function of $N$.
\begin{figure}[hbt]
\centerline{\strut\psfig{figure=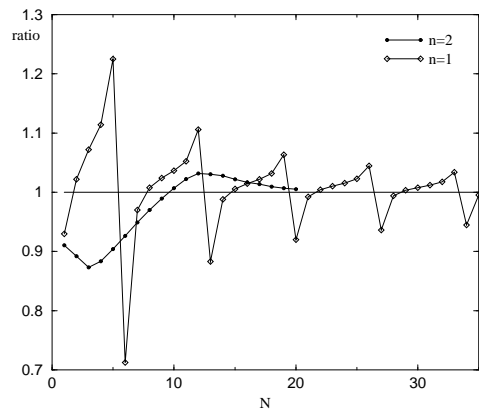,height=5.5cm}}
\caption{ The ratio of $(\mbox{Tr}\Lop^2)_N$ calculated via
the asymptotic formula ($\ref{result}$) to its value computed by 
numerical integration}
\label{abra}
\end{figure}

In summary we have  studied  the evolution operator for a discrete 
Langevin equation with a strongly hyperbolic 
classical dynamics and a Gaussian noise distribution. Using an 
integral representation of the evolution operator $\Lop$ we have 
revealed the asymptotic behaviour of the  corrections to 
the trace of $\Lop^n$. This behaviour is governed by sub-dominant
 terms corresponding 
to terms previously neglected in the perturbative expansion, and
 a fully analogous trace formula can be derived for the late
 terms in the noise extension series of the trace of $\Lop^n$.

G.V. and G.P. gratefully acknowledges the financial support of
the Hungarian Ministry of Education, EC Human Potential Programme, OTKA T25866.
G.P., G.V. and A.V. were also partially supported by the French Minist\`ere des
 Affaires {\'E}trang\`eres.


\end{document}